\documentclass[iop]{emulateapj}
\usepackage{amsmath}
\usepackage{graphicx}
\usepackage{color}
\usepackage{natbib}
\usepackage{soul}
\setcitestyle{authoryear,round,semicolon,aysep={,},yysep={,},notesep={; }}
\begin{document}
\title{The Star Formation Rate Function for Redshift $\MakeLowercase{z}\sim4-7$ Galaxies: Evidence for a Uniform Build-Up of Star-Forming Galaxies During the First 3 Gyr of Cosmic Time. }
\author{Renske Smit\altaffilmark{1}, Rychard J. Bouwens\altaffilmark{1,2}, Marijn Franx\altaffilmark{1},
 Garth D. Illingworth\altaffilmark{2}, Ivo Labb\'{e}\altaffilmark{1,4}, Pascal A. Oesch\altaffilmark{2,5},
 Pieter G. van Dokkum\altaffilmark{3}}
\altaffiltext{1}{Leiden Observatory, Leiden University, NL-2300 RA
 Leiden, Netherlands}
\altaffiltext{2}{UCO/Lick Observatory, University of California, Santa
 Cruz, CA 95064}
\altaffiltext{3}{Department of Astronomy, Yale University, New Haven, CT 06520}
\altaffiltext{4}{Carnegie Observatories, Pasadena, CA 91101}
\altaffiltext{5}{Hubble Fellow}

\begin{abstract}

We combine recent estimates of dust extinction at $z\sim4-7$ with UV luminosity function (LF) determinations to derive star formation rate (SFR) functions at $z\sim$4, 5, 6 and 7. SFR functions provide a more physical description of galaxy build-up at high redshift and allow for direct comparisons to SFRs at lower redshifts determined by a variety of techniques. Our SFR functions are derived from well-established $z\sim4-7$ UV LFs, UV-continuum slope trends with redshift and luminosity, and IRX-$\beta$ relations. They are well-described by Schechter relations. 
We extend the comparison baseline for SFR functions to $z\sim2$ by considering recent determinations of the H$\alpha$ and mid-IR luminosity functions. 
The low-end slopes of the SFR functions are flatter than for the UV LFs, $\Delta\alpha\sim+0.13$, and show no clear evolution with cosmic time ($z\sim0-7$).
In addition, we find that the characteristic value SFR$^\ast$ from the Schechter fit to SFR function exhibits consistent, and substantial, linear growth as a function of redshift from $\sim5\,M_\odot\rm{yr^{-1}}$ at $z\sim8$, 650 Myr after the Big Bang, to $\sim100\,M_\odot\rm{yr^{-1}}$ at $z\sim2$, $\sim2.5\,\rm{Gyr}$ later. 
Recent results at $z\sim10$, close to the onset of galaxy formation, are consistent with this trend. 
The uniformity of this evolution is even greater than seen in the UV LF over the redshift range $z\sim2-8$, providing validation for our dust corrections. 
These results provide strong evidence that galaxies build up uniformly over the first 3 Gyr of cosmic time.

\end{abstract}

\keywords{Galaxies: high-redshift --- Galaxies: evolution}

\section{Introduction}
In recent years great progress has been made in determining the UV luminosity function (LF) at early times \citep[e.g.,][]{Steidel1999,Wyder2005,Arnouts2005,ReddySteidel2009,Bouwens2007,Bouwens2011a,Oesch2010,Oesch2011}. Recent results on the UV LF indicate a bright end that builds up substantially with time, with the characteristic luminosity increasing from the highest redshifts to $z\sim3-4$ \citep[e.g.][]{Bouwens2007}. Furthermore, the faint-end slope of the UV LF is found to be very steep at redshifts $z\gtrsim3$ \citep[e.g.][]{Bouwens2011a}. This brightening of $L^\ast_{\rm{UV}}$ and steep faint-end slope $\alpha$ are consistent with a general picture where galaxies build up hierarchically and where lower luminosity galaxies play a major role in the reionization of the universe. 

Despite the general usefulness of the UV LF for probing early galaxy formation, one significant limitation of the UV LF in this regard is the sensitivity of rest-frame UV light to dust extinction. Inferring this attenuation directly at $z>3$ is challenging since only the most bolometrically luminous systems at these redshifts can be detected in far-IR. Also, for these very high redshifts, tracers of star formation such as X-ray, radio, 24$\mu$m and H$\alpha$ are either too faint to observe or redshifted out of the observable wavelength range of the most sensitive telescopes. 

Fortunately, we can take advantage of a relation between the UV-continuum slope $\beta$, with $f_\lambda\propto\lambda^\beta$ \citep{Meurer1999}, and the likely dust extinction to convert the observed UV-luminosities into SFRs \citep[see e.g.][]{Bouwens2009}. 
It has only now been possible to accurately establish the distribution of the UV-continuum slope distribution over a wide range in luminosity and redshift \citep[e.g.][]{Bouwens2009,Bouwens2011b,Wilkins2011,Finkelstein2011,Castellano2012,Dunlop2012}. This is the result of the installation of
the WFC3/IR camera on the Hubble Space Telescope and the recent acquisition of deep wide-area data over the HUDF09+CANDELS fields \citep{Bouwens2011b,Grogin2011,Koekemoer2011}. 

In this paper, we utilize these new UV-continuum slope determinations to derive SFR functions at $z\sim4$, 5, 6 and 7. SFR functions can better connect the growth of galaxies at the highest redshifts to galaxies found at later epochs, as well as give insight into the manner in which star formation lights up dark matter halos. Previously, such functions were presented at $z\sim0-1$ based on UV+IR data \citep{Martin2005,Bell2007,Bothwell2011}. We begin this paper by describing how we use $\beta$ to produce dust-corrected UV LFs and SFR functions (\S2). In \S3 we compare our SFR functions with the literature over a range of redshifts ($z\sim0-8$). For ease of comparison with previous studies we adopt $H_0=70\,\rm km\,s^{-1}\,Mpc^{-1},\,\Omega_{\rm{m}}=0.3\,$and$\,\Omega_\Lambda=0.7$. We adopt a \citet{Salpeter1955} IMF with limits 0.1-125$\,M_\odot$ throughout this paper.

\begin{figure}[t]
\centering
\includegraphics[width=0.9\columnwidth,trim=12mm 0mm 92mm 0mm]{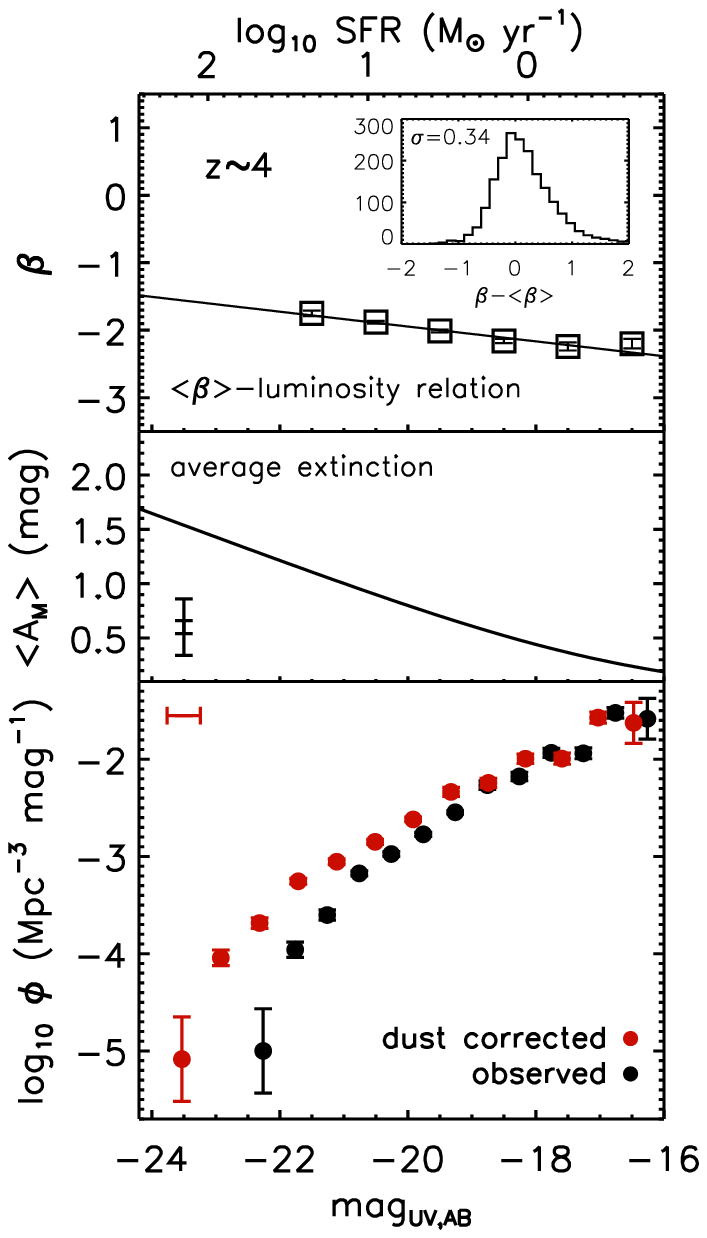}
\caption{\textit{Top}: The relation between the UV-continuum slope $\beta$ and UV luminosity at $z\sim4$ from \citet{Bouwens2011b} with the binned points and linear fit $<\beta>\,=-0.11\,(M_{\rm{UV,AB}}+19.5)-2.00$. The inset panel shows the luminosity de-trended distribution of UV-continuum slopes around the mean relation. \textit{Middle}: The average dust extinction from the $\beta$-distribution in the upper panel as a function of luminosity, assuming a \citet{Meurer1999} dust correction, as described in \S\ref{sec:dustcorr}. The errorbar in the bottom left corner shows both the random and systematic uncertainties in the relation. \textit{Bottom}: The black and red points show the $z\sim4$ UV LF from \citet{Bouwens2007} before and after correction for dust (see \S\ref{sec:dustcorr}). The errorbar in the top left corner shows the fiducial error in the average dust correction. The dust-corrected UV LF has a flatter faint-end slope $\alpha$ and brighter $M^\ast_{\rm{UV}}$.}
\label{fig:z4comparison}
\end{figure}

\begin{table}
\centering
\caption{Stepwise determinations of the SFR function at $z\sim4$, $z\sim5$, $z\sim6$ and $z\sim7$ (see \S\ref{sec:dustcorr})}

\begin{tabular}{lr}
\hline \hline 
log$_{10}\,\rm{SFR}\,(M_\odot\rm{yr^{-1}})$&$\rm\phi_{SFR}\,(Mpc^{-3}dex^{-1})$\\ 

\hline 
\multicolumn{2}{c}{$z\sim4$}\\
\hline
$-$0.66& 0.05920$\pm$0.02855\\
$-$0.44& 0.06703$\pm$0.00838\\
$-$0.21& 0.02537$\pm$0.00326\\
 0.02& 0.02534$\pm$0.00268\\
 0.25& 0.01430$\pm$0.00144\\
 0.48& 0.01153$\pm$0.00117\\
 0.72& 0.00601$\pm$0.00025\\
 0.95& 0.00354$\pm$0.00017\\
 1.19& 0.00221$\pm$0.00012\\
 1.44& 0.00139$\pm$0.00008\\
 1.68& 0.00052$\pm$0.00006\\
 1.92& 0.00023$\pm$0.00004\\
 2.16& 0.00002$\pm$0.00002\\
\hline
\multicolumn{2}{c}{$z\sim5$}\\
\hline
$-$0.33& 0.01766$\pm$0.00858\\
$-$0.11& 0.01161$\pm$0.00294\\
 0.12& 0.01076$\pm$0.00121\\
 0.36& 0.00420$\pm$0.00046\\
 0.61& 0.00362$\pm$0.00040\\
 0.86& 0.00224$\pm$0.00014\\
 1.11& 0.00121$\pm$0.00008\\
 1.37& 0.00060$\pm$0.00006\\
 1.63& 0.00023$\pm$0.00002\\
 1.89& 0.00006$\pm$0.00002\\
 \hline
\multicolumn{2}{c}{$z\sim6$}\\
\hline
$-$0.04& 0.01197$\pm$0.00262\\
 0.41& 0.00426$\pm$0.00089\\
 0.77& 0.00173$\pm$0.00037\\
 1.01& 0.00110$\pm$0.00024\\
 1.26& 0.00026$\pm$0.00008\\
 1.51& 0.00014$\pm$0.00004\\
 1.77& 0.00002$\pm$0.00002\\
 2.03& 0.00002$\pm$0.00002\\
\hline
\multicolumn{2}{c}{$z\sim7$}\\
\hline
$-$0.07& 0.01543$\pm$0.00473\\
 0.15& 0.00761$\pm$0.00215\\
 0.38& 0.00513$\pm$0.00149\\
 0.61& 0.00224$\pm$0.00075\\
 0.84& 0.00106$\pm$0.00044\\
 1.08& 0.00031$\pm$0.00019\\
 1.32& 0.00033$\pm$0.00018\\
\hline
\end{tabular} 
\label{tab:points}
\end{table}

\begin{figure*}
\centering
\includegraphics[width=0.85\textwidth,trim=5mm 35mm 0mm 10mm] {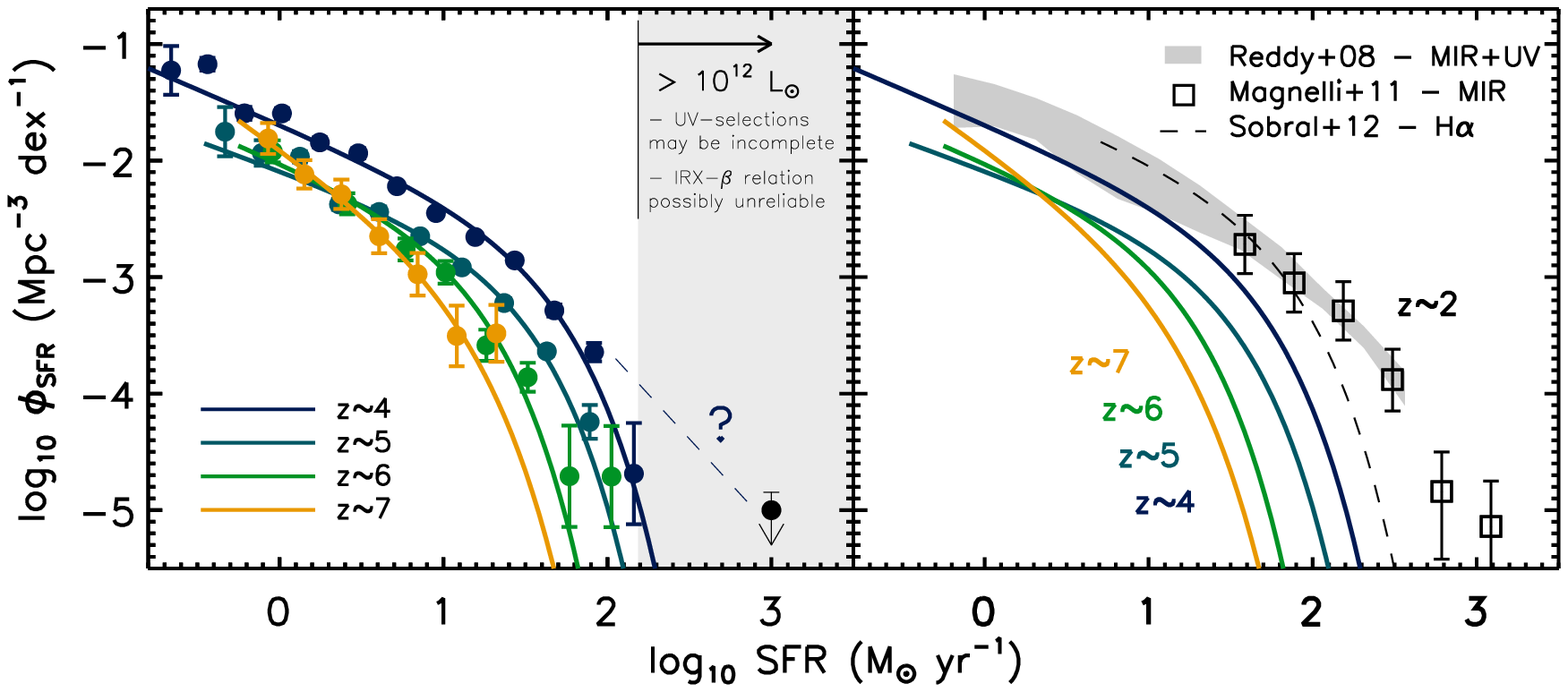} 
\caption{\textit{Left}: Both the analytical and stepwise SFR functions derived in this study from dust-corrected UV LFs. The stepwise SFR functions (individual points) were derived using the UV LF results from \citet{Bouwens2007,Bouwens2011a}, as described in \S\ref{sec:dustcorr}. The solid lines are the SFR functions derived in Schechter form as described in \S\ref{sec:analyticalSFR} with parameters listed in Table \ref{tab:SFRpar}. The lines are \textit{not} fits to the points. The excellent agreement between the two approaches provides a useful cross-check. 
We have indicated the SFRs ($>\sim150\,M_{\odot}\rm{yr^{-1}}$: equivalent to a bolometric luminosity $>10^{12}L_\odot$) where we expect our SFR functions to be more uncertain due to incompleteness in the UV selections and possible unreliability of the IRX-$\beta$ relation \citep[e.g.][]{Reddy2006}.  The best estimates at high SFRs may come from searches in the mid-IR/far-IR (black point from \citealp{Daddi2009}, see also \citealp{Marchesini2010}).   
The SFR function therefore may fall off more slowly than we infer (dashed line).
\textit{Right}: A comparison of the SFR functions with similar functions derived from the bolometric LF of \citet[\textit{grey region}]{Reddy2008}, the IR LF of \citet[\textit{black open squares}]{Magnelli2011} and the H$\alpha$ LF from \citet[\textit{black dashed line}]{Sobral2012} at $z\sim2$. The trend in the SFR function, derived from our dust-corrected UV LFs at $z\sim4-7$, clearly extends to $z\sim2$. The smooth evolution in the SFR function provides some corroboration for the dust corrections we apply. }
\label{fig:sfrfunc}
\end{figure*}

\begin{figure}
\centering
\includegraphics[width=0.9\columnwidth,trim=22mm 13mm 95mm -3mm]{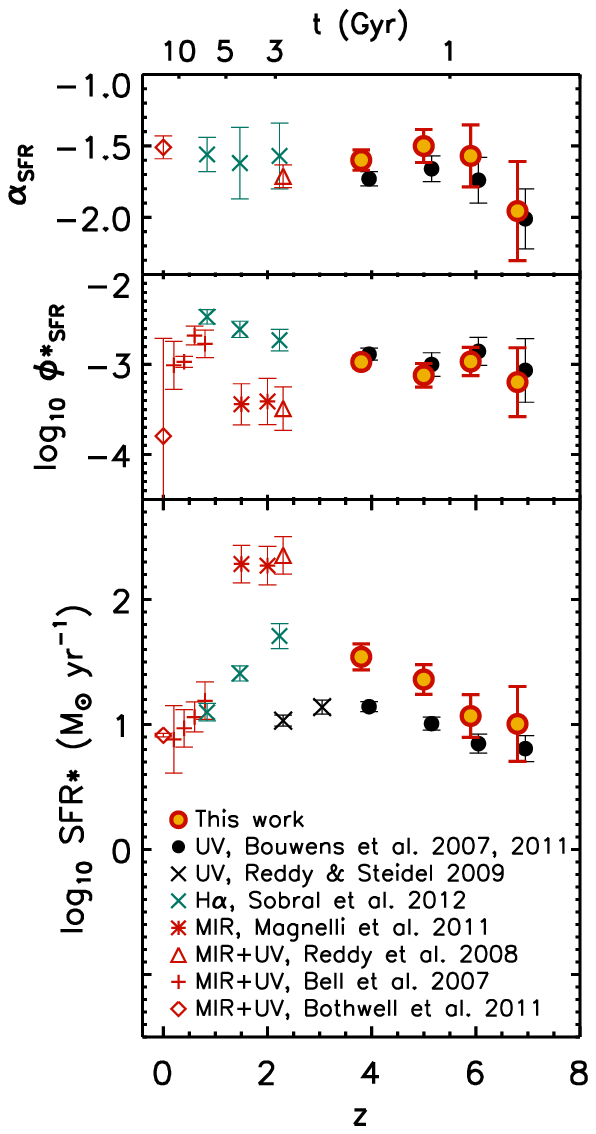}
\caption{Derived Schechter parameters $\alpha_{\rm{SFR}}$ (\textit{top}), $\phi_{\rm{SFR}}^\ast$ (\textit{middle}) and SFR$^\ast$ (\textit{bottom}) as a function of redshift. The red points are the derived dust-corrected SFR function parameters, while the black points and crosses show equivalent parameters for the UV LFs from \citet{Bouwens2007,Bouwens2011a} and \citet{ReddySteidel2009}, respectively, before dust correction (black points are shifted +0.15 on the x-axis for clarity). The errorbars for our parameters include both the random and systematic uncertainties in $\frac{d\beta}{dM}$
and $\beta_{\scriptstyle{M_{\rm{UV}}=-19.5}}$.
We compare to SFR parameters obtained from \citet[\textit{red diamonds}]{Bothwell2011}, \citet[\textit{red plus signs}]{Bell2007}, \citet[\textit{red triangles}]{Reddy2008}, \citet[\textit{red asterisks}]{Magnelli2011} and \citet[\textit{red crosses}]{Sobral2012}, as described in \S\ref{sec:results}. 
Our dust corrections result in flatter faint-end slopes $\alpha$ ($\Delta\alpha\sim+0.13$) relative to the UV LF (\S\ref{sec:faintend}). Our dust correction also doubles the rate at which $\rm{log_{10}\,SFR^\ast}$ grows with cosmic time over the redshift range $z\sim4-7$ (\S\ref{sec:brightend}). Clearly, dust corrections can have a significant impact on the apparent evolution of SF galaxies with time. }
\label{fig:var_z}
\end{figure}

\section{SFR functions}
This section describes how we compute SFR functions from published UV LFs and UV-continuum slopes $\beta$. We begin by deriving the SFR functions in a fully stepwise fashion, applying our luminosity-dependent dust corrections to binned UV LFs (Section 2.1). We then derive analytical formulae for these SFR functions based on the UV LFs we use as inputs (Section 2.2).

\subsection{Dust-corrected luminosity functions} 
\label{sec:dustcorr}

We correct UV LFs for the effects of dust attenuation using the well-known correlation of extinction with the UV-continuum slope $\beta$. 
We take the IRX-$\beta$ relation established by \citet{Meurer1999},
\begin{equation}
A_{\rm{1600}}=4.43+1.99\,\beta.
\label{eq:meurer}
\end{equation}
\citet{Meurer1999} estimated the relation based on starburst galaxies in the local universe. Similar relations have been found at $z\sim0$ by other groups \citep{Burgarella2005,Overzier2011}. Though there is some evidence that the \citet{Meurer1999} relation does not work well for all sources, e.g. very young galaxies ($<100$ Myr) and ULIRGs \citep[e.g.][]{Reddy2006}, this relation has been found to be accurate in the mean out to $z\sim2$, despite considerable scatter \citep[e.g.][]{Daddi2007,Reddy2006,Reddy2010,Reddy2012,Magdis2010a,Magdis2010b}. There is even evidence in the new Herschel observations that the \citet{Meurer1999} relation is reasonably accurate in estimating the dust extinction for $z\sim2$ Lyman-Break Galaxies \citep{Reddy2012}. We therefore quite reasonably utilize this relation in interpreting higher redshift samples.

Recently there have been a number of studies examining $\beta$ in high-redshift samples \citep[e.g.][]{Bouwens2009,Wilkins2011, Bouwens2011b,Finkelstein2011,Dunlop2012}. Perhaps the most definitive of these studies is \citet{Bouwens2011b}, who using the CANDELS+HUDF09 datasets find that $\beta$ correlates with both redshift and luminosity, with higher redshift and lower luminosity galaxies being bluer.
The results by \citet{Bouwens2011b} are in excellent agreement with other results in the literature \citep[e.g.][]{Ouchi2004,Labbe2007,Overzier2008,Wilkins2011}.

For our dust corrections we assume a linear relation between the UV-continuum slope $\beta$ and luminosity, such as that given in \citet{Bouwens2011b}. This is shown in the top panel of Figure \ref{fig:z4comparison} for $z\sim4$, 
\begin{equation}
<\beta>\,=\frac{d\beta}{dM_{\rm{UV}}}(M_{\rm{UV,AB}}+19.5)+\beta_{\scriptstyle{M_{\rm{UV}}=-19.5}},
\label{eq:beta}
\end{equation} 
where $\frac{d\beta}{dM_{\rm{UV}}}$ and $\beta_{\scriptstyle{M_{\rm{UV}}=-19.5}}$ are from Table 5 of \citet{Bouwens2011b}. Note that for $z\sim7$ we use a fit with a fixed slope $\frac{d\beta}{dM_{\rm{UV}}}=-0.13$ obtained from our $z\sim4-6$ samples, given the large uncertainties in this slope at $z\sim7$ and the lack of evidence for evolution in the $\beta$-luminosity relation over the redshift range $z\sim2-6$. 
The distribution of UV-continuum slope $\beta$ shows substantial scatter about relation 2, that can be approximated by a normal distribution (Figure \ref{fig:z4comparison}; but remark there is a fatter tail toward redder $\beta$).

The steps for computing an average dust correction are as follows. 
We use Eq. \ref{eq:meurer} to calculate the UV absorption $A_{1600}$ for each source in our adopted $\beta$-distribution. Then to obtain the extinction correction at a given $M_{\rm{UV}}$ we integrate over the $\beta$-distribution (normal distribution with mean $<\beta>$, Eq. \ref{eq:beta}, and $\sigma_\beta$=0.34), setting $A_{1600}=0$ when $A_{1600}<0$.
The middle panel of Figure \ref{fig:z4comparison} shows the resulting $<A_{\rm{M_{UV}}}>$ as a function of luminosity at $z\sim4$. We then apply this dust correction to the UV luminosities of individual bins of the LF. We shift each point in the LF toward brighter magnitudes and correct for the fact that the luminosity-dependent dust correction increases the width of the bins. 

The bottom panel of Figure \ref{fig:z4comparison} shows the effect of our luminosity-dependent dust correction on the stepwise UV LF at $z\sim4$ from \citet{Bouwens2007}. The dust correction shifts the LF to higher luminosities, particularly at the bright end, causing $M_{\rm{UV}}^\ast$ to brighten, and also flattens the faint-end slope. There is also a small shift to lower volume densities due to the renormalisation of the magnitude bins.

Now that we have dust-corrected UV fluxes we can use well-established relations to compute the SFR as a function of luminosity, giving us our desired SFR functions. We use the following relation from \citet{Kennicutt1998},

\begin{equation}
\frac{\rm{SFR}}{M_\odot\rm{yr^{-1}}}=1.25\cdot10^{-28}\,\frac{L_{\rm{UV,corr}}}{\rm{erg\,s^{-1}Hz^{-1}}}.
\label{eq:kennicutt}
\end{equation}
Since this relation gives the time-averaged SFR over a $\sim100\,\rm{Myr}$ time window, it will underestimate the SFR (typically by $\lesssim2\times$) in galaxies substantially younger than this \citep[e.g.][]{Verma2007,Bouwens2011b,Reddy2012}. However, Eq. \ref{eq:kennicutt} should work on average for the extended SF histories expected in LBGs.

The left panel of Figure \ref{fig:sfrfunc} shows the stepwise SFR functions at $z\sim4,\,5,\,6\,$and$\,7$, based on stepwise UV LFs derived from \citet{Bouwens2007} and \citet{Bouwens2011a}. For convenience, we include our stepwise SFR functions in Table \ref{tab:points}. 

\begin{table}
\centering
\caption{Schecher parameters determined for the present SFR functions}
\begin{tabular}{lccr}
\hline \hline 
$\langle z\rangle$&$\log_{10}\,\frac{\rm{SFR^\ast}}{M_\odot\,\rm{yr^{-1}}}$&$\log_{10}\,\frac{\phi_{\rm{SFR}}^\ast}{\rm{Mpc^{-3}}}$&$\alpha_{\rm{SFR}}$\\ 
\hline
3.8&1.54$\pm$0.10&1.07$\pm$0.17&$-$1.60$\pm$0.07\\
5.0&1.36$\pm$0.12&0.76$\pm$0.23&$-$1.50$\pm$0.12\\
5.9&1.07$\pm$0.17&1.08$\pm$0.39&$-$1.57$\pm$0.22\\
6.8&1.00$\pm$0.30&0.64$\pm$0.56&$-$1.96$\pm$0.35\\
\hline
\end{tabular} 
\label{tab:SFRpar}
\begin{flushleft}
\textbf{Notes.} These Schechter parameters are obtained by dust correcting the UV LF using the \citet{Meurer1999} IRX-$\beta$ relationship. We adopt the linear relation between the UV-continuum slope $\beta$ and UV luminosity (Eq. \ref{eq:beta}) found by \citet{Bouwens2011b}. See \S\ref{sec:analyticalSFR}.
\end{flushleft}
\end{table}

\begin{table*}
\centering
\caption{Schechter parameters for SFR functions in the literature (see also Figure \ref{fig:var_z})}
\begin{tabular}{lccccl}
\hline \hline 
$\langle z\rangle$&$\log_{10}\,\frac{\rm{SFR^\ast}}{M_\odot\,\rm{yr^{-1}}}$&$\log_{10}\,\frac{\phi_{\rm{SFR}}^\ast}{\rm{Mpc^{-3}}}$&$\alpha_{\rm{SFR}}$& probe&reference\\ 
\hline
0.0&0.91$\pm$0.01&$-$3.80$\pm$1.09&$-$1.51$\pm$0.08&MIR+UV&Bothwell et al. 2011$^a$\\
0.2&0.88$\pm$0.27&$-$3.01$\pm$0.27&$-$1.45(fixed)&MIR+UV&Bell et al. 2007\\
0.4&0.97$\pm$0.15&$-$2.97$\pm$0.06&$-$1.45(fixed)&MIR+UV&Bell et al. 2007\\
0.6&1.06$\pm$0.12&$-$2.68$\pm$0.10&$-$1.45(fixed)&MIR+UV&Bell et al. 2007\\
0.8&1.19$\pm$0.15&$-$2.77$\pm$0.15&$-$1.45(fixed)&MIR+UV&Bell et al. 2007\\
0.8&1.10$\pm$0.07&$-$2.47$\pm$0.08&$-$1.56$\pm$0.12&H$\alpha$&Sobral et al. 2012\\
1.5&1.41$\pm$0.06&$-$2.61$\pm$0.09&$-$1.62$\pm$0.25&H$\alpha$&Sobral et al. 2012\\
2.2&1.71$\pm$0.10&$-$2.73$\pm$0.12&$-$1.57$\pm$0.23&H$\alpha$&Sobral et al. 2012\\
1.5&2.28$\pm$0.15&$-$3.44$\pm$0.23&$-$1.60(fixed)&MIR&Magnelli et al. 2011\\
2.0&2.27$\pm$0.15&$-$3.41$\pm$0.26&$-$1.60(fixed)&MIR&Magnelli et al. 2011\\
2.3&2.35$\pm$0.15&$-$3.49$\pm$0.24&$-$1.71$\pm$0.08&MIR+UV&Reddy et al. 2008\\
3.8&1.54$\pm$ 0.10&$-$2.97$\pm$0.07&$-$1.60$\pm$0.07&UV+$\beta$&This work\\
5.0&1.36$\pm$ 0.12&$-$3.12$\pm$0.13&$-$1.50$\pm$0.12&UV+$\beta$&This work\\
5.9&1.07$\pm$ 0.17&$-$2.97$\pm$0.16&$-$1.57$\pm$0.22&UV+$\beta$&This work\\
6.8&1.00$\pm$ 0.30&$-$3.20$\pm$0.38&$-$1.96$\pm$0.35&UV+$\beta$&This work\\
\hline
\end{tabular} 
\label{tab:table3}
\begin{flushleft}
\textbf{Notes.} These Schechter parameters are derived from the published H$\alpha$, bolometric, and UV LFs in the literature using the \citet{Kennicutt1998} relations.

$^a$ It is unclear if there is a typographical error in the Schechter parameters provided by \citet{Bothwell2011}. A simple integration of \citet{Bothwell2011} SFR function does not yield the quoted SFR density. However we quote the numbers as they are presented in \citet{Bothwell2011}, converted to a Salpeter IMF with limits 0.1-125$\,M_\odot$.
\end{flushleft}
\end{table*}

\subsection{Analytical SFR functions} 
\label{sec:analyticalSFR}
In this section, we use an analytical Schechter-like approximation to represent SFR functions at $z\sim4-7$.

We assume that the IRX-$\beta$ relation for individual galaxies is described by $A_{\rm{UV}}=C_0+C_1\,\beta$ and the distribution of galaxies at a certain $M_{\rm{UV}}$ is given by a Gaussian with $\mu_\beta=<\beta>$ (Eq. \ref{eq:beta}) and width $\sigma_\beta$, which gives
\begin{equation}
<A_{\rm{M_{UV}}}>=C_0+0.2\ln10\,C_1^2\,\sigma_\beta^2+C_1<\beta>.
\label{eq:A}
\end{equation} 
This expression is only valid in the limit that the distribution of UV-continuum slopes $\beta$ does not extend to $\beta\lesssim-2.3$ since such blue $\beta$'s formally give negative dust corrections (a clearly non-physical result) using the \citet{Meurer1999} relation.
For the \citet{Meurer1999} relation Eq. \ref{eq:A} simplifies to $<A_{\rm{M_{UV}}}>=4.43+1.82\,\sigma_\beta^2+1.99<\beta>$.

To compute SFR functions we start with the UV LF, described in Schechter form \citep{Schechter1976}:
\begin{equation}
\phi(L)\,dL=\phi^\ast\,\left(\frac{L}{L^\ast}\right)^\alpha\,\exp\left(-\frac{L}{L^\ast}\right)\,\frac{dL}{L^\ast}. 
\end{equation}
Substituting SFR for L (and SFR$^\ast$ for L$^\ast$), using Eq. \ref{eq:beta}, \ref{eq:kennicutt} and \ref{eq:A} yields
 
\begin{eqnarray}\nonumber
\phi({{\rm{SFR}}})\,d{{\rm{SFR}}}=\frac{\phi^\ast}{1-C_1\,\frac{d\beta}{dM}}\,\left(\frac{{\rm{SFR}}}{{\rm{SFR}}^\ast}\right)^{\frac{\alpha+C_1\,\frac{d\beta}{dM}}{1-C_1\,\frac{d\beta}{dM}}}\\
\times\,\exp\left(-\frac{\rm{SFR}}{{\rm{SFR}}^\ast}\right)\,\frac{d{\rm{SFR}}}{{\rm{SFR}}^\ast}.
\end{eqnarray}
where we have made the simplifying assumption that the cut-off in the Schechter function is exponential and not some slightly shallower high-end cut-off (the modified functional form is consistent with the observations). This gives the conversions 
\begin{equation}
\label{eq:alpha}
\alpha_{\rm{SFR}}=\frac{\alpha_{\rm{UV,uncorr}}+C_1\,\frac{d\beta}{dM}}{1-C_1\,\frac{d\beta}{dM}}
\end{equation}
\begin{equation}
\label{eq:phi}
\phi_{\rm SFR}^\ast=\frac{\phi_{\rm{UV,uncorr}}^\ast}{1-C_1\,\frac{d\beta}{dM}}.
\end{equation} 

We calculate SFR$^\ast$ using Eq. \ref{eq:beta}, \ref{eq:kennicutt} and \ref{eq:A}, i.e. we use ${\rm{log_{10}\,SFR^\ast}}=-0.4\,(M^\ast_{\rm{UV,uncorr}}-<A_{\rm{M^\ast_{UV}}}>)-7.25$. Though Eq. \ref{eq:A} is a reasonable approximation for $<A_{\rm{M_{UV}}}>$, our estimate of SFR$^\ast$ is slightly more accurate when we estimate $<A_{\rm{M^\ast_{UV}}}>$ as described in \S\ref{sec:dustcorr}, by setting $A_{1600}=0$ when $A_{1600}<0$. Therefore we will use this more accurate estimate of SFR$^\ast$ quoted in Table \ref{tab:SFRpar}.

The left panel of Figure \ref{fig:sfrfunc} compares the analytical SFR functions (solid lines) with the corrected stepwise UV LFs described in \S\ref{sec:dustcorr}. They are in excellent agreement, providing a useful check 
on the analytic description used here. 
We would expect our derived SFR functions to be more uncertain at high bolometric luminosities ($L_{\rm{bol}}>10^{12}L_\odot$: indicated on Figure \ref{fig:sfrfunc})
where dust corrections are likely less reliable \citep[e.g.][]{Reddy2006} and our Lyman-break selections may be more incomplete to 
dusty star-forming galaxies \citep[e.g.][]{Daddi2009,Michalowski2010}. While this should not affect the turnover in the SFR 
function, the SFR function may fall off less steeply than the exponential form adopted here.

The analytical Schechter parameters are presented in Table \ref{tab:SFRpar}. The SFR$^\ast$, $\phi_{\rm{SFR}}^\ast$ and $\alpha_{\rm{SFR}}$ are calculated assuming the $z\sim4-7$ UV LF parameters from \citet{Bouwens2007} and \citet{Bouwens2011a}, the $\frac{d\beta}{dM}$, $\beta_{\scriptstyle{M_{\rm{UV}}=-19.5}}$ and $\sigma_\beta=0.34$ from \citet{Bouwens2011b} and the \citet{Meurer1999} relation (Eq. \ref{eq:meurer}). For the $z\sim7$ LF parameters, we modify our procedure somewhat due to the fact that Eq. \ref{eq:A} is not especially accurate for the very blue $\beta$'s found at $z\sim7$ (see discussion following Eq. \ref{eq:A}). Specifically, we keep the SFR$^\ast$ fixed to the value described above and then fit the low-end slope and the normalisation of the Schechter function to the stepwise UV LF.

We have verified through Monte-Carlo simulations that we can successfully recover ($\Delta{M}^\ast\lesssim0.13\rm\,mag,\,\Delta\alpha\lesssim0.03$) the observed UV LFs and $\beta$-luminosity relation using the derived SFR functions and a $\beta$-SFR relation \citep[e.g.][]{Castellano2012}, so our approach can be inverted. We describe these simulations in detail in Appendix A.

We remark that if $\beta$ shows a weaker dependence on luminosity than \citet{Bouwens2011b} find \citep[e.g.][]{Castellano2012,Dunlop2012} it would result in a steeper $\alpha_{\rm{SFR}}$ ($\Delta\alpha\sim0.1$) and higher $\phi_{\rm{SFR}}^\ast$ ($\Delta\phi_{\rm{SFR}}^\ast\sim0.05\,\rm{dex}$).
Uncertainties in both the incompleteness and contamination corrections used for the UV LFs, of course, also affect our SFR functions \citep[e.g.][]{Bouwens2007,Grazian2011}.

\begin{table}
\centering
\caption{Indicative SFRs over the redshift range $z\sim2$ to $z\sim8$}
\resizebox{1.0\columnwidth}{!}{
\begin{tabular}{lcccl}
\hline \hline 
$\langle z\rangle$&{ $\log_{10}\,\frac{\rm{SFR^\ast}}{M_\odot\,\rm{yr^{-1}}}$}&{ $\log_{10}\,\frac{\rm{SFR}}{M_\odot\,\rm{yr^{-1}}}$}& { probe}&{ reference}\\ 
&{(fixed $\phi^\ast$)}$^a$&{ (fixed $ n(\rm >SFR)$)$^b$}&&\\ 
\hline
2.0&2.04$\pm$0.08&2.01$\pm$0.30$^c$&  { MIR}& Magnelli et al. 2011\\
2.2&1.89$\pm$0.02&1.80$\pm$0.30$^c$&{H$\alpha$}&Sobral et al. 2012\\
2.3&2.08$\pm$0.03&2.05$\pm$0.30$^c$&{ MIR+UV}&Reddy et al. 2008\\
3.8&1.58$\pm$0.10&1.52$\pm$0.30$^c$&{ UV+$\beta$}&This work\\
5.0&1.33$\pm$0.11&1.27$\pm$0.30$^c$&{ UV+$\beta$}&This work\\
5.9&1.09$\pm$0.16&1.05$\pm$0.30$^c$&{UV+$\beta$}&This work\\
6.8&0.93$\pm$0.29&0.86$\pm$0.30$^c$&{ UV+$\beta$}&This work\\
8.0&0.61$\pm$0.04&0.49$\pm$0.30$^c$&{ UV}&Oesch et al. 2012a\\
\hline
\end{tabular} }
\label{tab:table4}
\begin{flushleft}
\textbf{Notes.} These Schechter parameters are derived from the published H$\alpha$, bolometric luminosity and UV luminosity functions in the literature using the \citet{Kennicutt1998} relations.

$^a$ Fixed $\rm{log_{10}\,\phi^\ast_{SFR}}=-3.05$. See also Figure \ref{fig:sfrstar_z}.

$^b$ Fixed number density $n(\rm >SFR)=2\cdot10^{-4} Mpc^{-3}$ \citep{Papovich2011}.

$^c$ Indicative errors from \citet{Papovich2011}. If we use a \citet{Chabrier2003} IMF as in \citet{Papovich2011} (instead of a \citealt{Salpeter1955} IMF with limits 0.1-125$\,M_\odot$) the SFRs given here would be 0.2 dex lower.
\end{flushleft}
\end{table}

\begin{figure*}
\centering
\includegraphics[width=0.85\textwidth,trim=0mm 0mm 15mm 40mm]{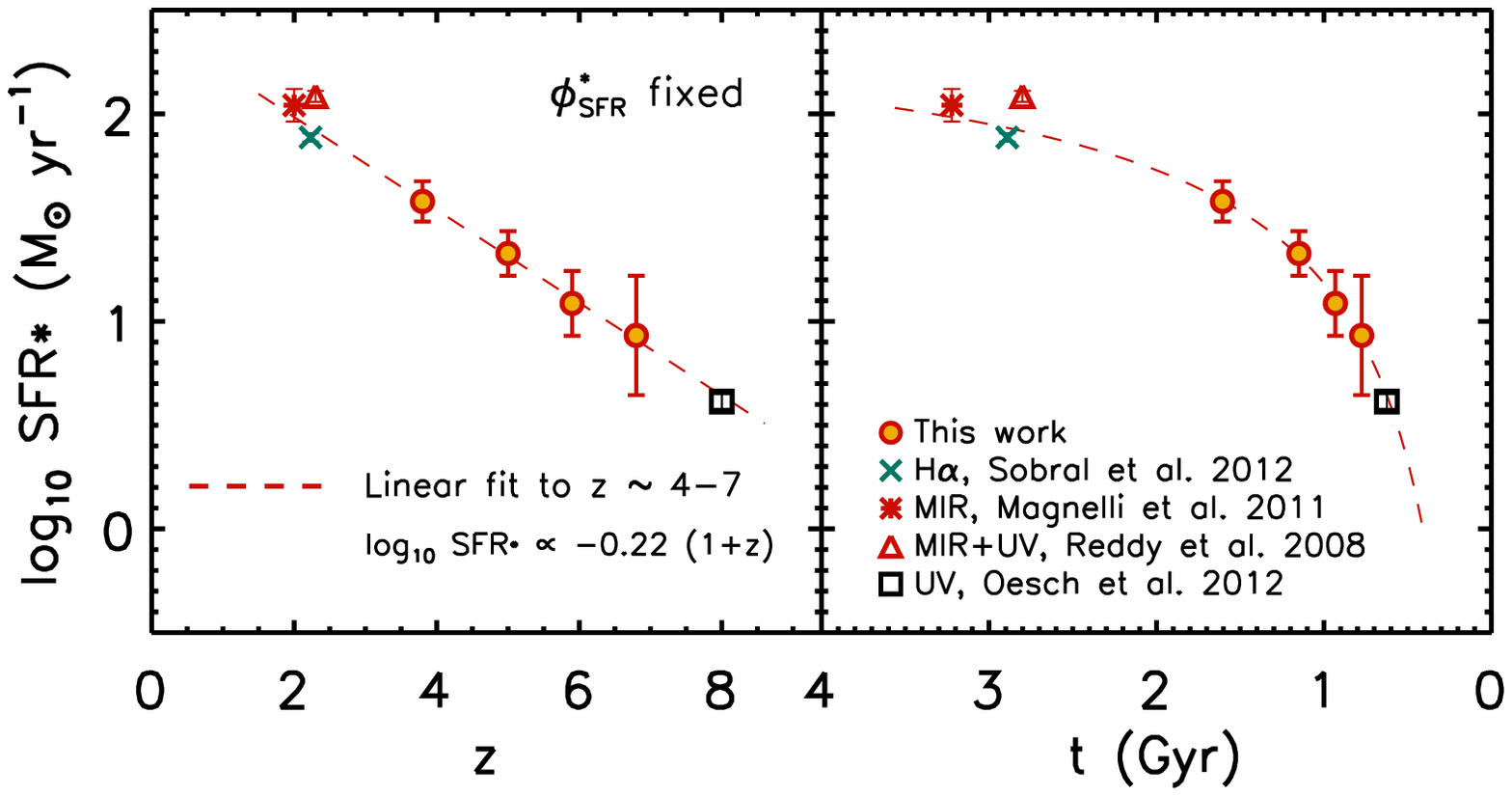}
\caption{The turnover SFR$^\ast$ at fixed $\phi_{\rm{SFR}}^\ast$ as a function of redshift (\textit{left}) and cosmic time (\textit{right}). The SFR$^\ast$ from \citet{Magnelli2011}, \citet{Sobral2012}, \citet{Reddy2008} and \citet{Oesch2012} are indicated with a red asterisk, cross, triangle and square, respectively. The dashed line gives a linear fit to our derived $\rm{log_{10}\,SFR^\ast}$ at $z\sim4-7$, with $\rm{log_{10}\,SFR^\ast\propto-0.22\,(1+z)}$ (\S\ref{sec:brightend}). 
The fit at $z\sim4-7$ is in broad agreement with the literature over the entire range $z\sim8$ to $z\sim2$. The extrapolated fit also agrees with the current estimate for SFR$^\ast$ at $z\sim10$ \citep{Bouwens2011nature,Oesch2011}, but the uncertainties in this estimate are still quite large. This provides strong evidence that galaxies build up consistently with time, from $z\sim8$ to $z\sim2$.}
\label{fig:sfrstar_z}
\end{figure*}

\section{Results}
\label{sec:results}
One of the main results of this paper is our determination of the approximate Schechter parameters for the SFR function. These parameters allow for an intuitive and quantitative look at how this function evolves with cosmic time, as well as allowing for a comparison with lower redshift determinations. We discuss both the high-SFR and low-SFR end of the SFR function.

\subsection{High-SFR end: linear build-up of log SFR$^\ast$ from $z\sim8$ to $z\sim2$}
\label{sec:brightend}
The high-SFR end of the SFR function is interesting since it provides us with a direct measure of the rate at which luminous galaxies are building up at early times.  
We compare our parameters with various studies that combine UV and IR observations to obtain complete SFR functions at $z\lesssim2$, i.e. \citet{Bothwell2011}, \citet{Bell2007} and \citet{Reddy2008} and also with H$\alpha$ and IR LFs at $z\sim2$ from \citet{Sobral2012} and \citet{Magnelli2011} respectively. We expect the IR LFs to probe the SFRs in the dominant population of dusty galaxies (ULIRG+LIRGs) at  $z\sim2$.
We use the relations from \citet{Kennicutt1998} to convert the H$\alpha$ and bolometric luminosities to SFRs. 
The SFR functions at $z\sim2$ are shown in the right panel of Figure \ref{fig:sfrfunc}; these SFR functions are in reasonable agreement with each other, except at the high-SFR end where the H$\alpha$ LF is low. This may result from an incomplete sampling (or inadequate dust corrections) of dusty galaxies by the H$\alpha$ study.
The $z\sim2$ SFR functions are consistent with the evolution observed between $z\sim4-7$.

The bottom panel of Figure \ref{fig:var_z} (see also Table \ref{tab:table3}) shows SFR$^\ast$ for our results at $z\sim4-7$ with the SFR$^\ast$ from published studies mentioned above. 
The black symbols \citep{Bouwens2007,Bouwens2011a,ReddySteidel2009} represent the equivalent SFR$^\ast$ one would derive without applying a dust correction. 
There is a significant difference in the evolution of the high-SFR end of the SFR function taking dust attenuation into account: SFR$^\ast$ (without a dust correction) peaks at $\sim20\,M_\odot\rm{yr^{-1}}$ for $z\sim3-4$, while SFR$^\ast$ derived from dust-corrected UV or MIR LFs continues to increase from $\sim10\,M_\odot\rm{yr^{-1}}$ at $z\sim7$ to $\sim100\,M_\odot\rm{yr^{-1}}$ at $z\sim2$.

In Figure \ref{fig:sfrstar_z} we show the evolution of SFR$^\ast$ at fixed $\phi_{\rm{SFR}}^\ast$ (with $\rm{log_{10}\,\phi^\ast_{SFR}}=-3.05$: see also Table \ref{tab:table4}). This is interesting since it allows us to examine the evolution in the high-end of the SFR function without introducing additional "noise" from the SFR$^\ast$-$\phi_{\rm{SFR}}^\ast$ degeneracy. We extend our comparison of SFR$^\ast$ to $z\sim8$ using the \citet{Oesch2012} UV LF \citep[see also][]{Trenti2011,Bradley2012}. The effect of the dust correction on SFR$^\ast$ decreases strongly with redshift for $z\gtrsim4$ and we expect the contribution at $z\sim8$ to be nearly negligible.
We fit a linear slope to our own estimates of $\rm{log_{10}\,SFR^\ast}$ at $z\sim4-7$ and find $\rm{log_{10}\,SFR^\ast=2.43-0.22\,(1+z)}$ (Figure \ref{fig:sfrstar_z}).   
The SFR functions without a dust correction follow $\rm{log_{10}\,SFR^\ast\propto-0.11\,(1+z)}$. 
Comparing the two, we see that our dust corrections double the rate at which the high-SFR end of the SFR function grows with cosmic time (see also the discussion in \S6.2 of \citealt{Bouwens2006}).

Extrapolating the best fit SFR$^\ast$-redshift relation to higher and lower redshift we find good agreement with $z\sim2$ and $z\sim8$ determinations of SFR$^\ast$ (Figure \ref{fig:sfrstar_z}).  
The estimate for $z\sim10$ from \citet{Oesch2011} is also consistent with the trend, but the uncertainty in SFR$^\ast$ at that redshift is still quite large. The evolution of SFR$^\ast$ with redshift at fixed $\phi^\ast$ suggests that galaxies build up in a consistent way during the first 3 Gyr of cosmic time. Moreover, the excellent agreement between different probes of star formation across cosmic time, provides a valuable cross-check on the validity of the dust corrections we use.

\citet{Papovich2011} also consider an evolving SFR vs. redshift relation based on an abundance-matched selection. Not surprisingly, given that \citet{Papovich2011} use essentially the same UV LFs \citep[from][]{Bouwens2007,Bouwens2011a} and similar dust corrections at $z\gtrsim4$ as we use here (from \citealt{Bouwens2009} instead of \citealt{Bouwens2011b}), they find similar SFRs over the range $z\sim4-8$ (at constant number density $n(\rm >SFR)=2\times10^{-4}\rm{Mpc}^{-3}$, see Table \ref{tab:table4}). However, the present analysis suggests much more clearly that the build-up in the SFRs of galaxies extends not simply from $z\sim8$ to $z\sim4$, but continues all the way to $z\sim2$.

\begin{figure*}[t]
\centering
\includegraphics[width=0.85\textwidth]{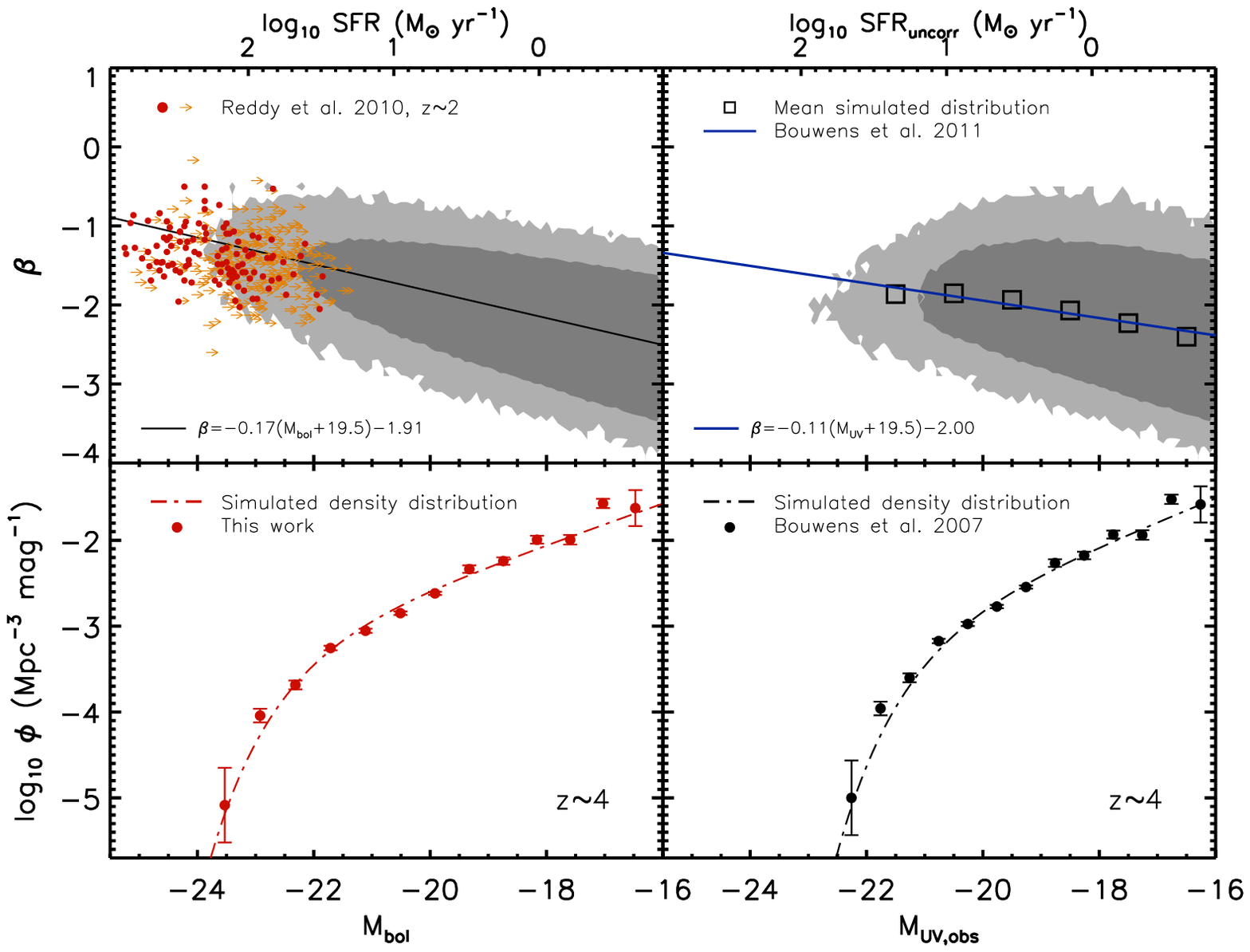}
\caption{Demonstration that we can use the SFR function to recover the two inputs to the SFR function as described in Appendix A. We show the bolometric luminosity function at $z\sim4$ (equivalent to the SFR function with parameters in Table \ref{tab:SFRpar}) in the lower left panel and a $\beta$ vs. bolometric luminosity relation (equivalent to a $\beta$ vs. SFR relation) in the upper left panel. The upper right panel shows the $\beta$ vs. observed UV luminosity relation we are attempting to reproduce with our simulations, and the lower right panel shows the UV LF. The gray area in the top left panel indicates our initial mock dataset, with each mock galaxy having a $\beta$ drawn from a normal distribution with a mean $<\beta>\,=-0.17\,(M_{\rm{bol}}+19.5)-1.91$ and a $L_{\rm{bol}}$ from the bolometric LF in the bottom left panel. Also plotted are the bolometric luminosities (\textit{red dots}) and upper limits (\textit{orange arrows}) of \citet{Reddy2010} based on their rest-frame UV-luminosities and \textit{Spitzer} MIPS 24$\mu\rm{m}$ flux measurements. The top right panel shows the simulated distribution as a function of UV luminosity, which we derive by dust correcting the bolometric luminosities based on the intrinsic $\beta$ values for the individual mock galaxies. The black squares indicate the mean $\beta$'s we recover as a function of luminosity in these simulations, while the blue line gives the observed relation found in \citet{Bouwens2011b}. The bottom right panel shows that the UV LF we recover from these simulations (dot dashed lines) matches up well with the observed UV LF from \citet{Bouwens2007} at $z\sim4$ (black circles and error bars).
}
\label{fig:appendix}
\end{figure*}

\subsection{Low-SFR end: flatter slopes and large uncertainties}
\label{sec:faintend}
The low-end slope of the SFR function is interesting, since it provides us with insight into the physics and feedback effects in the lowest luminosity galaxies when compared to the dark matter halo mass function.
The top panel of Figure \ref{fig:var_z} shows how the low-end slope $\alpha$ of the SFR function evolves as a function of redshift. 
The evolution of dust-corrected $\alpha_{\rm{SFR}}$ contrasts with the original slope of the UV LF. $\alpha_{\rm{SFR}}$ is typically $\Delta\alpha\sim+0.13$ flatter on average. Comparing $\alpha_{\rm{SFR}}$ with similar determinations at lower redshift \citep{Bothwell2011,Reddy2008,Sobral2012}, we find no evidence for evolution with cosmic time.

\section{Summary}
In this paper we combine UV LFs with dust corrections based on the UV-continuum slope $\beta$ and the IRX-$\beta$ relation, to produce star formation rate functions between $z\sim 4-7$. The dust correction results in a flatter faint-end slope and brighter $M^\ast_{\rm{UV}}$.

We find SFR functions that, at fixed $\phi^\ast$, show a steady build-up in their value of SFR$^\ast$ with cosmic time, from $\sim5\,M_\odot\rm{yr^{-1}}$ at $z\sim8$ to $\sim100\,M_\odot\rm{yr^{-1}}$ at $z\sim2$, providing evidence that galaxies build up in a very uniform fashion in the first 3 Gyr (to $z\sim2$).  Use of the SFR function allows us to naturally link the evolution of normal star-forming galaxies at $z\gtrsim4$ with the population of dusty star-forming galaxies seen at $z\sim2-3$.

\acknowledgments
We thank Naveen Reddy, David Sobral and Marcel van Daalen for useful discussions. We are thankful to Naveen Reddy for sending us the information necessary to make a comparison with the bright $z\sim2$ spectroscopic sample from \citet{Reddy2010}. Feedback from the referee significantly improved this paper. We acknowledge the support of a NWO vrije competitie grant, ERC grant HIGHZ \#227749, and NASA grant HST-GO-11563.01.  PO acknowledges support from NASA through a Hubble Fellowship grant \#51278.01 awarded by STScI.

\section*{Appendix}
\section*{A. Monte-Carlo simulation to recover the UV LF and $\beta$-luminosity relation}
In this study we derive SFR functions by transforming the UV LF using a luminosity dependent dust correction. However, one might suppose that the dust correction may be better parametrized in terms of the SFR instead of the UV luminosity -- given that the SFR is the more physical quantity and dust extinction is known to be well-correlated with the SFR \citep[e.g.][]{Reddy2006}. It is therefore worthwhile investigating whether we can start with our derived SFR functions and then recover the two inputs to the SFR function, i.e., the rest-frame UV LF and the observed $\beta$-UV luminosity relationship. This will allow us to test whether the SFR functions we derive are sensitive to our parametrizing the mean $\beta$ as a function of observed UV luminosity (instead of parametrizing it in terms of the SFR). We perform this test at $z\sim4$, where both the observed UV-luminosity function and the observed $\beta$-UV luminosity relation are best determined.

The steps in the simulation are as follows. We start with our $z\sim4$ SFR function and we produce a mock data set with each galaxy in the data set having a SFR and $\beta$. The initial $\beta$ distribution is drawn from a normal distribution with a mean value equal to $<\beta>\,=\frac{d\beta}{dM_{\rm{bol}}}(M_{\rm{bol}}+19.5)+\beta_{\scriptstyle{M_{\rm{bol}}=-19.5}}$ and a scatter equal to $\sigma_{\rm{int}}$. This is shown with the grey area in the top left panel of Figure \ref{fig:appendix} as $\beta$ versus bolometric luminosity and the simulated bolometric luminosity function is shown on the bottom left panel of Figure \ref{fig:appendix}.

Assuming the \citet{Meurer1999} relation to calculate dust attenuation, we obtain the observed UV luminosities for each galaxy in this mock data set from the SFRs, as shown in the top right panel of Figure \ref{fig:appendix}. We derive the $\beta$ versus SFR distribution in an iterative process so as to best reproduce the observed $\beta$-UV luminosity relationship with $<\beta>\,=-0.11\,(M_{\rm{UV,AB}}+19.5)-2.00$ and observed scatter $\sigma_{\rm{obs}}=0.34$ \citep{Bouwens2011b}. The best-fit $\beta$ vs. $M_{\rm bol}$ distribution has the form: $<\beta>\,=-0.17\,(M_{\rm{bol}}+19.5)-1.91$ and $\sigma_{\rm{int}}=0.30$.

As a check on our derived $\beta$ vs. SFR distribution we have plotted the measurements from \citet{Reddy2010} in the top left panel of Figure \ref{fig:appendix}. \citet{Reddy2010} use rest-frame UV and 24$\mu\rm{m}$ observations of $z\sim2$ galaxies to study the properties of star-forming galaxies as a function of their bolometric luminosities. The agreement between the bright end of our simulated dataset and the \citet{Reddy2010} measurements is encouraging, though we note that at fixed bolometric luminosity one might expect $z\sim2$ galaxies to have somewhat higher dust content and therefore redder $\beta$ values than our $z\sim4$ galaxies due to evolution in dust extinction vs. SFR relation \citep[e.g.][]{Reddy2006,Buat2007}. The \citet{Reddy2010} measurments therefore provide a rough upper limit to the steepness of the $\beta$-SFR relation at $z\sim4$. 

We convert the simulated observed distribution to a luminosity function and compare this simulated observed UV LF with the UV LF of \citet[see the bottom right panel of Figure \ref{fig:appendix}]{Bouwens2007}. The two LFs are in quite reasonable agreement. The characteristic luminosity M$^\ast$ we find is within 0.13 mag of the derived LF of \citet{Bouwens2007} and the faint-end slope $\alpha$ we find is within 0.03 of the slope \citet{Bouwens2007} find. These differences are smaller than our current error bars for our SFR function parameters at $z\sim4$.\footnote[1]{We also considered the possibility that our results are affected by scatter in the IRX-$\beta$ relation at $z\gtrsim4$. For example if we assume that dust extinction can be derived from the observed $\beta$ distribution and this $\beta$ distribution is a simple function of the SFR (as discussed in Appendix A), we estimate that a 0.2 dex scatter in the IRX-$\beta$ relation could possibly result in a somewhat lower value of SFR$^\ast$ ($\rm\Delta{SFR}^\ast\sim0.09\rm\,dex$). However, we could just as easily have assumed that dust extinction can be derived from the SFR of a galaxy and that scatter in the $\beta$ distribution at a given SFR is due to scatter in the IRX-$\beta$ relation which would result in a somewhat higher value of SFR$^\ast$ (which is in the opposite sense from the previously discussed scenario). On balance, we cannot correct our results for the effects of scatter in the IRX-$\beta$ relation unless we have detailed knowledge about how dust extinction, $\beta$, UV luminosity and SFR are correlated. In any case it would appear that any possible corrections are small, $\sim 0.1$ dex, and one must be encouraged by how well our SFR function results match up with the LF results at $z\sim2$.}

This simulation gives us confidence that our derived SFRs are not subject to large biases due to incompleteness at the faint end of the distribution. Furthermore, these simulations suggest that dust extinction is not so large at the bright end of the LF that we are unable to approximately recover the SFR function. Overall, these test results provide some validation for our methodology for performing the dust corrections.

\bibliographystyle{apj}

\end{document}